\documentclass[fleqn,usenatbib]{mnras}

\usepackage{newtxtext,newtxmath}

\usepackage[T1]{fontenc}

\DeclareRobustCommand{\VAN}[3]{#2}
\let\VANthebibliography\thebibliography
\def\thebibliography{\DeclareRobustCommand{\VAN}[3]{##3}\VANthebibliography}

\usepackage{graphicx}	%
\usepackage{amsmath}	%
\usepackage{siunitx}  %
\usepackage{subfig}

\newcommand{\rms}[1]{\ensuremath{_{\text{#1}}}}

\newcommand{\tsu}[1]{\textsubscript{#1}}
\newcommand{\atom}[2]{$^{#2}\text{#1}$}
\newcommand{\weblink}[1]{\texttt{\href{#1}{#1}}}

\newcommand{\elvis}{\texttt{I2ELVIS}}
\newcommand{\al}{\atom{Al}{26}}
\newcommand{\fe}{\atom{Fe}{60}}
\newcommand{\water}{{H\tsu{2}O}}

\newcommand{\waterfraci}{f_{\text{H}_2\text{O},i}}
\newcommand{\waterfract}{f_{\text{H}_2\text{O},t}}
\newcommand{\waterfracf}{f_{\text{H}_2\text{O},f}}

\newcommand{\waterfracfinal}{\waterfracf/\waterfraci}
\newcommand{\waterfracH}{\mathcal{H}}
\newcommand{\waterfracHf}{\mathcal{H}_f}

\title[Devolatilization of planetesimals by SLRs]{Devolatilization of extrasolar planetesimals by \fe{} and \al{} heating}

\author[J. W. Eatson et al.]{
Joseph W. Eatson$^{1}$\thanks{E-mail: \texttt{\href{mailto:j.w.eatson@sheffield.ac.uk}{j.w.eatson@sheffield.ac.uk}}},
Tim Lichtenberg$^{2}$,
Richard J. Parker$^{1}$\thanks{Royal Society Dorothy Hodgkin Fellow}
\&
Taras V. Gerya$^{3}$
\\
$^{1}$Department of Physics and Astronomy, The University of Sheffield, Hicks Building, Hounsfield Road, Sheffield, S3 7RH, UK\\
$^{2}$Kapteyn Astronomical Institute, University of Groningen, P.O. Box 800, 9700 AV Groningen, NL\\
$^{3}$Institute of Geophysics, Department of Earth Sciences, ETH Zurich, Sonneggstrasse 5, 8092 Zurich, CH\\
}

\pubyear{2023}

\begin{document}
\label{firstpage}
\pagerange{\pageref{firstpage}--\pageref{lastpage}}
\maketitle

\begin{abstract}
\noindent
Whilst the formation of Solar system planets is constrained by meteoritic evidence, the geophysical history of low-mass exoplanets is much less clear. The bulk composition and climate states of rocky exoplanets may vary significantly based on the composition and properties of the planetesimals they form from. An important factor influenced by planetesimal composition is water content, where the desiccation of accreting planetesimals impacts the final water content of the resultant planets. While the inner planets of the Solar system are comparatively water-poor, recent observational evidence from exoplanet bulk densities and planetary formation models suggest that rocky exoplanets engulfed by substantial layers of high-pressure ices or massive steam atmospheres could be widespread. Here we quantify variations in planetesimal desiccation due to potential fractionation of the two short-lived radioisotopes \al{} and \fe{} relevant for internal heating on planetary formation timescales. We focus on how order of magnitude variations in \fe{} can affect the water content of planetesimals, and how this may alter the formation of extrasolar ocean worlds. We find that heating by \al{} is the dominant cause of planetesimal heating in any Solar system analogue scenario, thus validating previous works focussing only on this radioisotope. However, \fe{} can become the primary heating source in the case of high levels of supernova enrichment in massive star-forming regions. These diverging scenarios can affect the formation pathways, bulk volatile budget, and climate diversity of low-mass exoplanets.
\end{abstract}


\section{Introduction}

Astronomical observations \citep{2022MNRAS.511.4551L,2022Sci...377.1211L,2022AJ....164..172D,2023NatAs...7..206P} and formation models \citep{2019PNAS..116.9723Z,2020A&A...643L...1V} have recently opened new frontiers in characterising low-mass exoplanets, highlighting a potential deviation between volatile-rich and under-dense versus volatile-poor and mostly rocky worlds \citep{2021ApJ...922L...4D,Zilinskas2023,2023arXiv230610100P}. In the early Solar System, short-lived radioisotopes (SLRs) sensitively affected the thermal evolution, differentiation, and volatile composition of planetesimals and forming planets \citep{2017pedc.book..115F,2018SSRv..214...39M,2023ASPC..534..907L,2022arXiv220310056K}, which can shape system-to-system deviations in exoplanet volatile trends \citep{lichtenbergWaterBudgetDichotomy2019,2021ApJ...913L..20L}.

Substantial internal heating of planetesimals by SLRs, in particular \al{}, in the Solar System was first theorised by \cite{ureyCosmicAbundancesPotassium1955}.
Evidence of a substantial \al{} enrichment in the early Solar System in the form of an excess of the \al{} decay product \atom{Mg}{26} in calcium–aluminium-rich inclusions (CAIs) was discovered by \citet{grayExcess26MgAllende1974} and \citet{1976GeoRL...3...41L}.
With such levels of enrichment, \al{} was shown to be a dominant heating mechanism for planetesimals and asteroidal precursor bodies \citep{1989Icar...82..244G,grimmHeliocentricZoningAsteroid1993}.  
The excess levels of \fe{} have been less clear from meteoritic data \citep{2012E&PSL.359..248T,2018GeCoA.221..342T,2018ApJ...857L..15T,2021ApJ...917...59C}, with estimates varying over two orders of magnitude (from $^{60}\mathrm{Fe}/^{56}\mathrm{Fe} = 10^{-8}$ to $10^{-6}$) in recent years, with the lower \fe{} levels favoured in most recent measurements. At the lower end limit, the abundances of \fe{} are not expected to be sufficient to cause substantial heating and volatile loss of planetesimals, however, substantial variations of the \al{}/\fe{} ratio are expected across star-forming regions \citep{2013ApJ...769L...8V,2019ApJ...870....3B,2022ApJ...933....1B,2022ApJ...925...56F,2023MNRAS.521.4838P}.

Multiple formation mechanisms exist for both \al{} and \fe{} \citep{2018PrPNP.102....1L,2021PASA...38...62D,2023JPhG...50c3002L}.
While spallation from stellar cosmic rays has been suggested as a potential mechanism for enrichment of \al{} \citep{2021ApJ...919...10A}, the low levels of $^{10}$Be \citep{2000Sci...289.1334M,2022GeCoA.324..194D} and mostly homogenous distribution of decay products throughout the Solar System \citep{2022arXiv220311169D} suggest that this mechanism is not the most influential \citep{parkerBirthEnvironmentPlanetary2020}. 
Mechanisms external to a stellar system are more probable, such as through supernovae \citep{chevalierYoungCircumstellarDisks2000,2010ARA&A..48...47A,2010ApJ...711..597O,2014MNRAS.437..946P,2015PhyS...90f8001P}, stellar winds \citep{2009ApJ...696.1854G}, or AGB stars \citep{2018PrPNP.102....1L}.
Supernovae in particular produce both \fe{} and \al{} in abundance, with usually elevated \fe{}/\al{} ratios relative to the Solar System \citep{2016ApJ...826...22K,lichtenbergIsotopicEnrichmentForming2016}, while stellar winds from massive evolved stars (such as Wolf-Rayets) can produce significant quantities of \al{} over their lifetime \citep{limongiNucleosynthesis26Al60Fe2006,limongiPresupernovaEvolutionExplosive2018}.
As the distribution of almost all SLRs are approximately homogenous, the half life times are short compared to the lifetime of the protoplanetary disks, and \al{}-devoid CAIs (FUN-CAIs) are rare \citep{villeneuveHomogeneousDistribution26Al2009,2023Icar..40215611D}, it is likely that SLRs are deposited immediately prior to or during the formation of the first CAIs.
As such, we can infer that the enrichment of SLRs in a stellar system is sensitive to the local stellar environment immediately leading up to the formation of the stellar system \citep{2010ARA&A..48...47A,2019A&A...622A..69P,parkerBirthEnvironmentPlanetary2020}.

Heating from SLRs has a significant impact on the volatile evolution of initially ice-rich planetesimals \citep{2017pedc.book...92C,2021Icar..36314437G,2023Natur.615..854N}, which can have a knock-on effect on the properties of subsequent planets \citep{grimmHeliocentricZoningAsteroid1993,2021Sci...371..365L}.
Of importance for this paper, planetesimal internal heating from the decay of SLRs can result in melting, heating, vaporisation and subsequent outgassing of volatiles -- particularly water and volatile carbon compounds.
In the case of stellar systems with significant SLR enrichment, this can result in a significant reduction in surface and internal water content, which is a prime marker for bulk composition and affects the long-term climatic evolution of rocky planets. Constraining the influence of radiogenic heating for multiple radioisotopes on the volatile loss from planetesimals during accretion would allow for improvements in planetary formation models, better estimations of the ocean world population, and quantify the possible fractionation between major atmophile elements across planetary populations \citep{2019MNRAS.482.2222W,2023ApJ...948...53S}.

Previous simulations of planetesimal devolatilization primarily focused on the influence of \al{} on planetesimal water and carbon compounds \citep{lichtenbergWaterBudgetDichotomy2019,2021ApJ...913L..20L,2022ApJ...938L...3L}. This work, in contrast, specifically focusses on the influence of \fe{} and thus varying \fe{}/\al{} ratios on planetesimal desiccation.
A parameter space exploration is performed to determine how \fe{} heating can affect planetesimals of varying sizes, iron contents and degree of \fe{} enrichment relative to the Solar System.
Internal core-mantle differentiation and thus \fe{} distribution is simulated by use of two distinct internal structure models, which are discussed in Section \ref{sec:meth}.
As there are multiple parameters to be explored, the number of simulations was extremely large, as such, 2D models with a temperature-based rather than a chemistry-based model were utilised.
Additionally, constraining some of these parameters for subsequent simulation sets was the basis for a number of simulation subsets, which are described in Section \ref{sec:results}.
Finally, the results are discussed and conclusions are made in Section \ref{sec:discuss} \& \ref{sec:conclusion}.

\section{Methodology}
\label{sec:meth}

\subsection{Numerical \& heating simulation}

\begin{figure}
    \centering
    \subfloat[A comparison of retained water fraction over time for a number of simulations of different resolutions.]{
        \includegraphics[scale=0.65]{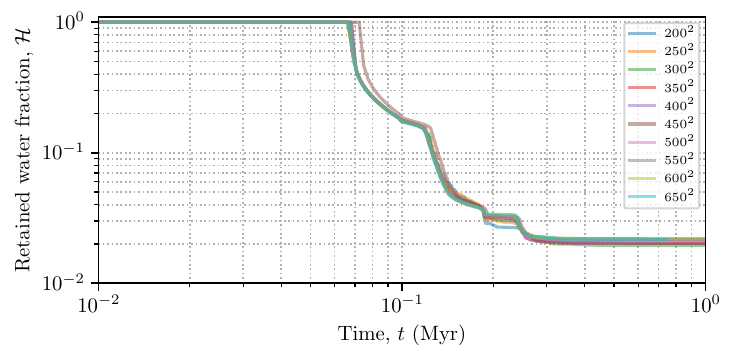}
    }
    
    \subfloat[A comparison of final water fraction over simulation cell count. The final result is insensitive to simulation resolution.]{
        \includegraphics[scale=0.65]{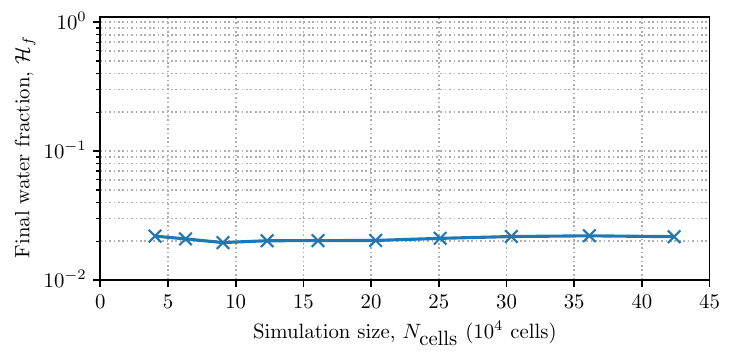}
    }
    \caption{A comparison of \water{} desiccation for a set of simulations with varying resolution. These simulations were conducted to determine whether simulation resolution impacted the results of the simulations. We found that there was no significant change as resolution varies, as such a value of $250^2$ cells was utilised as a good compromise between speed and accuracy.}
    \label{fig:resolutions}
\end{figure}

Numerical simulations of planetesimals were conducted using the \elvis{} thermo-mechanical geodynamic code\footnote{\url{https://github.com/FormingWorlds/i2elvis_planet}}, which uses a combination of the marker-in-cell and conservative finite difference schemes to simulate planetesimal evolution \citep{geryaCharacteristicsbasedMarkerincellMethod2003a,2007PEPI..163...83G}, with a similar setup as in \citet{lichtenbergWaterBudgetDichotomy2019} and \citet{2021Sci...371..365L}. 
Simulations were conducted with a fixed grid resolution of $250 \times 250$ cells.
Higher resolutions were tested, but found to produce similar results at the expense of significantly increased processing time (Fig. \ref{fig:resolutions}).
Simulations evolve using a variable timestep, which is constrained by the following parameters:

\begin{itemize}
  \item \texttt{maxxystep}: The maximum displacement distance for material in a cell ($\partial x / \partial t \cdot \Delta t$).
  \item \texttt{maxtkstep}: The maximum temperature change per step.
  \item \texttt{maxtmstep}: The maximum total timestep.
\end{itemize}

\noindent
Additionally, the maximum timestep is set to $2500$ years throughout the simulation, in order to prevent numerical instability.

Upon initialisation \elvis{} assigns properties to cells in order to make up the planetesimal and its surrounding environment.
These properties are analogous to geological compositions within the planetesimal:%

\begin{itemize}
  \item ``Sticky air'', the surrounding medium that ensures thermal stability, a proxy for the medium surrounding the planetesimal \citep{crameriComparisonNumericalSurface2012}.
  \item Hydrous silicates which contain water and the simulation's supply of \al{}.
  \item Iron, which contains the simulation's supply of \fe{}.
\end{itemize}

Hydrous silicates have three important states in a simulation that determine the water content of the planetesimal.
Initially water in the wet silicates are assumed to be stored as ice, which can undergo phase changes based on temperature.
Beyond a temperature of \SI{273}{K} ($T\rms{melt}$) the ice melts, which delays further temperature increase as the enthalpy of fusion needs to be overcome.
This phase change is reversible, and re-freezing can occur if the cell falls below the threshold temperature.
Beyond a temperature of \SI{1223}{K}, the upper limit of the amphibolite stability field, the entire water content of the cell becomes totally and irreversibly vaporised, and is assumed to leave the planetesimal through out-gassing processes (though these are not simulated); this temperature is referred to as $T\rms{vap}$.
Water retention across the planetesimal is stored as a fraction of the form:

\begin{equation}
  \waterfracH = \frac{\waterfract}{\waterfraci}\equiv \frac{N\rms{wet}}{N\rms{dry}} , 
\end{equation}

\noindent
where $\waterfract$ is the volume of hydrous silicates that has never been heated above $T\rms{vap}$, $\waterfraci$ is the initial volume of hydrous silicates, $N\rms{wet}$ is the number of hydrous silicate cells still containing water in the liquid or ice phases, and $N\rms{dry}$ is the number of hydrous silicate cells that have been removed of water.
This ratio is calculated and stored for every timestep.

An important variable discussed is the final retained water fraction, $\waterfracHf$, which is the value of $\waterfracH$ at the end of the simulation ($\waterfracHf = \waterfracf / \waterfraci$).
Measuring the exact volume of remaining water introduces a significant number of additional parameters, and as such is beyond the scope of this project.
This method can be performed for a number of other chemicals, such as CH$_4$, CO$_2$, or CO \citep{2021ApJ...913L..20L}, but for simplicity we focus here on water as the main volatile.

Heating of the planetesimal is provided by the decay of \al{} and \fe{} isotopes.
The specific heating of an SLR is given by the formula:

\begin{equation}
  \label{eq:specificheating}
  H\rms{SLR}(t) = \frac{N\rms{A} E\rms{SLR} \lambda}{m\rms{SLR}} e^{-\lambda t},
\end{equation}

\noindent
where $N\rms{A}$ is Avogadro's constant\footnote{\num{6.02214076e23}}, $E\rms{SLR}$ is the SLR decay energy, $\lambda$ is the decay constant of the SLR ($\lambda = \ln(2) / \tau\rms{1/2}$) and $m\rms{SLR}$ is the molar mass of the SLR.
The specific heating rate for SLRs embedded within a rocky body can be calculated with the equation:

\begin{equation}
  \label{eq:heatingrate}
  Q\rms{SLR}(t) = f\rms{E,CI} Z\rms{SLR} H\rms{SLR}(t) ,
\end{equation}

\noindent
where $f\rms{E}$ is the chrondritic elemental abundance and $Z\rms{SLR}$ is the radioisotopic enrichment \citep{castillo-rogez26AlDecayHeat2009}.
Enrichment is measured relative to the early Solar System estimates of \fe{} and \al{} enrichment; the Solar System enrichment values for \al{} and \fe{} are detailed in Table \ref{tab:isotopes}.
The total heating rate can then be calculated through the formula:

\begin{equation}
  \label{eq:heatingbody}
  Q\rms{T}(t) = Q\rms{26Al}(t) + Q\rms{60Fe}(t) ,
\end{equation}

\noindent
where $Q\rms{26Al}$ and $Q\rms{60Fe}$ are the individual heating contributions per unit mass of \al{} and \fe{}.

\begin{figure}
    \centering
    \subfloat[Specific heating, based on Equation \ref{eq:heatingrate}.]{
        \includegraphics[scale=0.65]{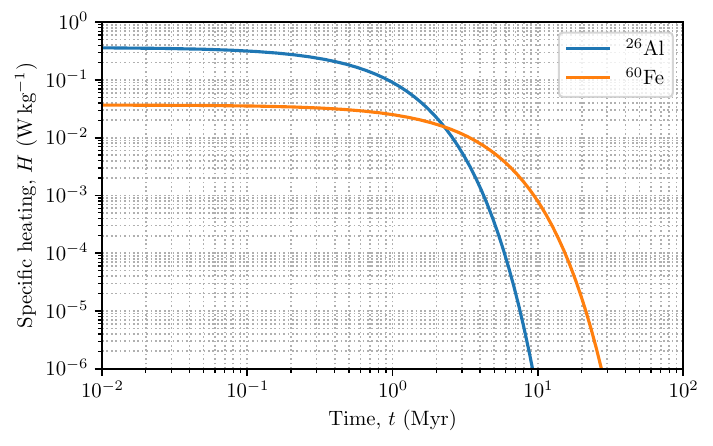}
    }
    
    \subfloat[Planetary body specific heating rate, based on Equation \ref{eq:heatingbody}.]{
        \includegraphics[scale=0.65]{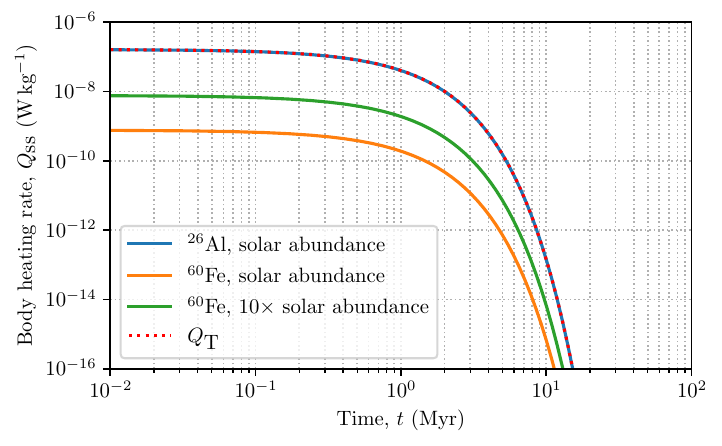}
    }
        
    \caption{A comparison of the specific heating, $H\rms{SLR}$, and specific heating rates, $Q\rms{SLR}$, of the SLRs \al{} and \fe{} at Solar System abundances and enrichments. Whilst the specific heating rate of \fe{} remains higher for longer, Solar System abundances result in a minimal impact on the total heating rate. The tables are produced using Eq. \ref{eq:specificheating} and Eq. \ref{eq:heatingrate} using parameters detailed in Table \ref{tab:isotopes}.}
    \label{fig:heating}
\end{figure}

Fig. \ref{fig:heating} shows decay heating from \al{} and \fe{} sources over time, as well as specific heating rates for the SLRs with Solar System abundances.
As can be seen, \al{} specific heating is significantly higher for the first 2 million years after CAI formation, however this is supplanted by \fe{} after this point.
In the case of Solar System abundances, however, \al{} heating dominates due to the lower quantity of \fe{} over the entire period of planetesimal formation and cooling.
In other star systems that had interactions with supernovae during their formation we may find a greater abundance of \fe{}, which would impact radiogenic heating and the water budget of subsequently forming planets.

Throughout this paper the SLR enrichment is described relative to the estimated isotopic enrichment of the Solar System after CAI formation.
This is parametrised as the SLR enrichment ratio, $\Lambda$.

\begin{equation}
  \Lambda\rms{SLR} = \frac{Z_{\text{SLR},\star}}{Z_{\text{SLR},\odot}} , 
\end{equation}

\noindent
where $Z_{\text{SLR},\star}$ is the post-CAI formation SLR enrichment of the system described in the simulation and $Z_{\text{SLR},\odot}$ is the equivalent for the Solar System.
Table \ref{tab:isotopes} notes the enrichment values for each  for Solar System estimates.
Simulations are run for a total of \SI{10}{Myr}, we assumed that the planetesimal being simulated is formed at \SI{1}{Myr} after formation of the first CAIs.
As such \SI{1}{Myr} of radioactive decay is simulated prior to the start of hydrodynamical simulations to reflect this.

\begin{table}
  \centering
  \begin{tabular}{ccccc}
    \hline
    Isotope & $f$ & $\tau$ & $E$ & $Z_{\text{SLR},\odot}$ \\
    \hline
    \al{} & 0.0085 & \SI{0.717}{Myr} & \SI{3.210}{MeV} & \num{5.250e-5}\\
    \fe{} & 0.1828 & \SI{2.600}{Myr} & \SI{2.712}{MeV} & \num{1.150e-8} \\
    \hline
  \end{tabular}
  \caption{SLR properties, with the associated Solar System SLR-to-stable isotope ratios \citep{2013M&PS...48.1383K,2012E&PSL.359..248T}.}
  \label{tab:isotopes}
\end{table}

\subsection{Assumptions \& limitations}
\label{sec:assumptions}

Our simulations are based on static planetesimals that do not accrete further, meaning that ongoing accretion of, for example, varying composition of pebbles that grow planetesimals is not taken into account \citep{2021Sci...371..365L,2023A&A...671A..74J}. The addition of accreting pebble layers on top of planetesimals has been shown to alter their thermal evolution \citep{2022Icar..38515100S}. Furthermore, we do not model geochemical reactions operating in the host rock, such as the oxidation of iron metal by water flow. This is an important aspect, as the diversity of FeO and hydrogen-bearing phases in inner Solar System asteroids bear evidence of rapid water-rock reactions during planetary formation \citep{2017RSPTA.37560209S,2018SSRv..214...39M,2019E&PSL.52615771M,grant2023bulk}, and may influence the final volatile fractionation on planetesimals. Following up from this work, we will thus investigate the more detailed multi-phase dynamics of water-rock aggregates \citep{2005E&PSL.240..234T,2017SciA....3E2514B}. This is a substantial simplification as mobilization and transport of volatile ices, fluids, and gases strongly influences their redistribution between forming metal core, planetary mantle, and outgassed reservoir \citep{suer2023_frontearthsci}. Additionally, the physical dislocation in multi-phase fluid treatments can differ substantially if the initial aggregate consists of a mixture that becomes comparable in ice and rock composition \citep{gerya2019introduction}. \citet{2023ApJ...956L..25Z} recently suggested sublimation of volatile ice phases as an additional devolatilization mechanism for ice-rich planetesimals that cross to the inside of the water snow line \citep[due to the snow line migrating outwards during the Class I stage of the solar protoplanetary disk,][]{2021Sci...371..365L}, further highlighting the importance of multiphasic treatments of planetesimals. In a similar vein, ongoing metal-silicate differentiation is not treated directly, but bracketed by our assumptions on Fe distribution, which we discuss in the next subsection.

\subsection{\fe{} distribution models}

Shortly after the formation of the stellar system, subsequent rapid formation of planetesimals, and initial radogenic heating, the hot, undifferentiated liquidus phase material of the planetesimal begins to undergo differentiation \citep{2015GMS...212...83N}.
Segregation of silicates and iron occur on an approximately similar timescale, which occur over a period of $0.4\,\si{Myr}$ to $10\,\si{Myr}$ \citep{2019E&PSL.507..154L}, with the upper limit being the approximate timescale of our simulation.
The segregation time of the planetesimal material is dependent on the initial formation time, degree of radiogenic heating \citep{doddsThermalEvolutionPlanetesimals2021}, and mode of core formation \citep{2020E&PSL.54616419W}.
As \al{} heating drives melting of silicates, a strongly convective mantle forms, which can inhibit the growth of the core \citep{neumannDifferentiationCoreFormation2012}.
As the planetesimal cools this convective flow decreases and eventually the mantle solidifies, this occurs on a timescale of $10\,\si{Myr}$ to $50\,\si{Myr}$ \citep{neumannMultistageCoreFormation2018}.
There are two qualitative types of core formation in planetesimals, which differ in timescale: in percolative core formation, S-rich metals may form an interconnected vain network, which can remove mantle metals quickyl via gravitational drainage before the appearance of the first silicate melts \citep{2003Natur.422..154Y,2017PNAS..11413406G}. However, laboratory experiments find that typically some metals remain stranded in the silicate matrix \citep{2009E&PSL.288...84B,2015E&PSL.417...67C,2023E&PSL.61718247W}, which are then removed from the mantle once the silicates reach the rheological transition.
As the range of core formation times across these two major modes intersects the epoch our simulations occur in, and a full model of core formation and material segregation are beyond the scope of this work, two distinct planetesimal structures are simulated in this paper:

\begin{enumerate}
  \item The \emph{core model}: Iron is contained in a large metal core surrounded by silicates. This represents an idealised end-state for a pre-differentiated planetesimal, in a scenario where metal-silicate separation operates rapidly and early.
  \item The \emph{grain model}: Iron is randomly distributed throughout the planetesimal. This represents the end-member scenario of metal core formation completing slowly, after the main stage of internal devolatilization.
\end{enumerate}
    
\begin{figure}
  \centering
  \includegraphics[scale=0.7]{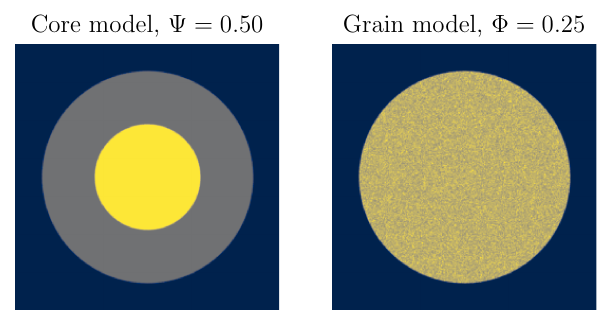}
  \caption{Comparison of the initial conditions of each model with an approximately equivalent iron fraction of $N\rms{Fe}/N\rms{T} = 0.25$. The \emph{core model} distributes all iron (yellow) into a central core surrounded by silicates (grey), with a planetesimal radius ratio of $\Psi$ (Eq. \ref{eq:psi}). The \emph{grain model} randomly distributes iron throughout the planetesimal with a cell fraction of $\Phi$ (Eq. \ref{eq:phi}).}
  \label{fig:distromodels}
\end{figure}

\noindent 
Fig. \ref{fig:distromodels} illustrates the physical differences of these models, as well as their model-dependent parameters.
An additional model with a mantle containing iron grains as well as a small, still forming core was considered, as it represented the mid-point between the two models.
However, simulations with these models were not conducted as the results from the first two models were very similar, as discussed in Section \ref{sec:grainmodelresults}.
In addition to the parameters dictating iron abundance in each model, $\Psi$ and $\Phi$, the planetesimal radius and isotopic enrichment are varied, in order to explore a broad parameter space.

\subsubsection{Core model} %
\label{sec:coremodel}

\noindent
The first model considered for this research consists of an iron core surrounded by hydrous silicates.
This model is comparatively easy to implement, and required no modification of the underlying \elvis{} code.
However, this model is not a realistic case for planetesimal formation while still retaining water, as iron core formation by liquid iron alloy percolation requires heating of a planetesimal above the dehydration temperature of \SI{1223}{K} to \SI{1700}{K} \citep{neumannDifferentiationCoreFormation2012}.
Whilst this case therefore does not represent and accurate account of how metal and volatile parts of a planetesimal would behave, it can approximate SLR repartitioning during core formation, which occurs in tandem with dehydration.
As such, it represents an idealised end-member case, and can be compared with the successor grain model to determine how different distributions of iron and its related \fe{} heating throughout the planetesimal body may affect the final water fraction.

The size of the core is controlled by the core radius ratio parameter, given by the equation:

\begin{equation}
  \label{eq:psi}
  \Psi = \frac{r\rms{c}}{r\rms{pl}} , 
\end{equation}

\noindent
where $r\rms{c}$ is the core radius and $r\rms{pl}$ is the total planetesimal radius.
Throughout these simulations this parameter is varied from 0.0 up to 0.99; values beyond this were considered redundant as this is already an extreme value, and that there would be no hydrous silicate cells left to measure desiccation from.

Whilst simple to execute, this model is less physically accurate than the \emph{grain model} also described in this paper.
Primarily, the degree of differentiation for even large planetesimals is comparatively slow, and occurs over a timescale of $\sim 1 \, \si{Myr}$ to $\sim 10 \, \si{Myr}$ after CAI formation.
As such, a differentiated body is unlikely for a large planetesimal so early after its formation.
However, with more extreme degrees of radiogenic heating from \fe{} the process of iron melt segregation may be accelerated due to faster melting of the body.

\subsubsection{Grain model}

The \emph{grain model} provides a more realistic simulation of the planetesimal shortly after formation.
Upon initialisation a percentage of the cells containing hydrous silicates are converted at random to have an iron composition instead.
The model therefore produces planetesimals where core formation has not occurred.
A pseudorandom number generator is used to determine which cells are converted, as true random numbers were not deemed necessary.

As the percentage of cells affected by the change increases the amount of iron in the planetesimal increases, we define the Fe grain volume ratio, $\Phi$, as the ratio of the number of cells changed at initialisation to the number of grains that were unaffected:

\begin{equation}
  \label{eq:phi}
  \Phi = \frac{N\rms{Fe}}{N\rms{S}} , 
\end{equation}

\noindent
where $N\rms{Fe}$ is the number of cells with the iron marker and $N\rms{S}$ is the number of cells with the hydrous silicate marker upon initialisation.
This parameter is used as a stand-in for the iron content in the planetesimal; similarly to the \emph{core model}, \fe{} enrichment is a separate parameter.

This model offers significant improvements over the iron \emph{core model} in terms of physical accuracy, as it offers a closer analogue to the post-formation conditions and properties of a typical planetesimal.
Additionally, planetesimals under these conditions should evenly heat faster than planetesimals with an iron core, due to the even distribution of \fe{} throughout the rocky body of the planetesimal.
However, whilst more accurate than the iron \emph{core model}, the \emph{grain model} still has some physical inaccuracies.
In particular, iron and silicates cannot segregate due to limitations in the one-phase fluid approximation, though below the iron alloy melting temperature this is not a concern \citep{2019GeoJI.219..185K,2021JGRE..12606754Z}.

\subsection{Data recording \& processing}

2D outputs are performed every 50 timesteps, and contain the primitive variables (density, velocity \& pressure), temperature and radiogenic heating rate.
These outputs can then be used to calculate other values and can be averaged for specific regions of the planetesimal.
Additionally, hydrous fraction, ice fraction and water fraction are dumped for the entire planetesimal at every time step.

\section{Results}
\label{sec:results}

\begin{figure}
    \centering
    \includegraphics[scale=0.7]{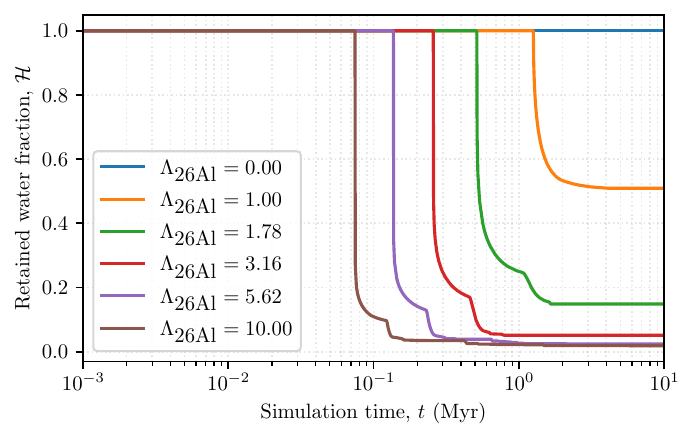}
    \caption{\emph{Core model} comparison of water fraction, $\mathcal{H}$, as a function of simulation time for planetesimals of radius \SI{50}{km} and an un-enriched iron core with a radius ratio of $0.25$. Simulations follow a similar structure, with most desiccation occurring over a very short period, before reaching a minimum as radiogenic heating reduces.}
    \label{fig:timecomparison}
\end{figure}

For simulations where desiccation occurs we found that the evolution of the planetesimals was broadly the same.
As the planetesimal begins to heat up over the first $10^5\,\si{yr}$ of the simulation some cells exceed the melting temperature, before subsequently exceeding the vaporisation temperature.
Desiccation occurs rapidly at this point, before the planetesimal begins to thermalise, then cool, preventing any further desiccation from occurring.
This progression can be seen in Fig. \ref{fig:timecomparison}, where this radpid desiccation can be observed.

The parameter space exploration of this paper was broken into a series of simulations.
Firstly, the \emph{core model} was utilised in order to get a baseline of how \fe{} heating affects water content, and its relative heating impact compared to \al{}.
This model was also used to constrain the ideal size of planetesimal for subsequent simulations, and also to determine how iron content can affect the efficiency of \al{} as a heating element.
Subsequently, the \emph{grain model} was used for a separate set of simulations, in order to determine how iron distribution affects heating, in order to determine if heating is the only major variable in desiccation.

\subsection{Core model}

\subsubsection{Core size ratio}

First, we performed a set of simulations in order to explore how core size -- and therefore iron abundance -- affects desiccation of planetesimals.
The core-to-radius ratio, $\Psi$, was varied between 0.05 and 0.95 in steps of 0.05, with additional simulations of 0.01 and 0.99 for completeness.
Simulations for $\Psi = 0$ and $\Psi = 1$ were not conducted as these were determined to be redundant (Section \ref{sec:coremodel}).
The common parameters for these simulations were a radius of \SI{100}{km} with \fe{} enrichment a factor of $10^3$ greater than Solar System estimates.
No \al{} enrichment was included in these simulations, in order to focus solely on desiccation due to \fe{} heating.
The results of this set of simulations were then used to constrain the core size parameter space, in order to reduce the number of required simulations.

\begin{figure}
  \centering
  \includegraphics[scale=0.65]{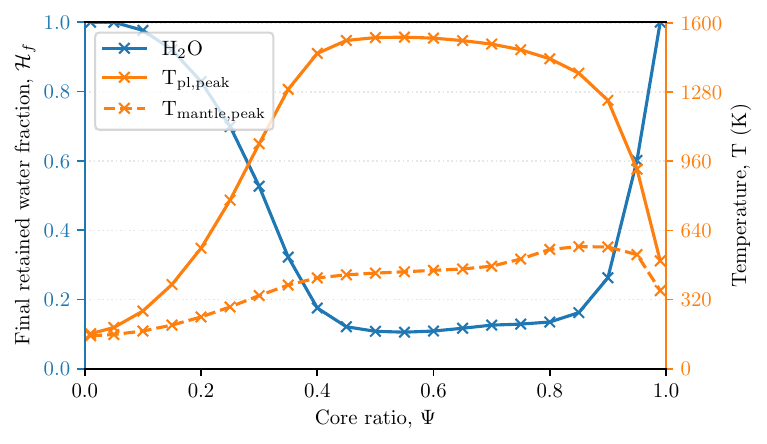}
  \caption{\emph{Core model} simulation of final retained \water{} fraction, $\waterfracfinal$, peak mean planetesimal temperature and peak mean mantle temperature for simulations varying core size, $\Psi$, with \fe{} enrichment $10^3$ times greater than Solar System estimates. Desiccation becomes significant after the core radius, $r\rms{c}$, is greater than 20\% of the radius of the planetesimal, $r\rms{p}$, quickly reaches a maxima and decreases rapidly beyond 90\% due to only crustal deposits remaining.}
  \label{fig:core-ratios}
\end{figure}

Fig. \ref{fig:core-ratios} shows that initially desiccation is fairly limited, but rapidly increases as the core-to-planetesimal ratio increases.
Eventually a point is reached where increasing the core size does not lead to any appreciable increases in desiccation amount.
Desiccation increases rapidly when the planetesimal is $\sim 1\%$ iron by volume, plateaus at $\sim 10\%$, and decreases again at $\gtrsim 50\%$ iron by volume.
As a result, two values of $\Psi$ were chosen for further simulations:

\begin{itemize}
  \item $\Psi = 0.25$, which is in the range of the initial increase in desiccation, and corresponds to a Fe volume fraction of $1.56\%$.
  \item $\Psi = 0.50$, which is in the trough of maximum desiccation, and corresponds to a Fe volume fraction of $12.5\%$.
\end{itemize}

\subsubsection{Radius comparison}

After determining how desiccation amount varies through modification of the core size, we then progressed to determining how desiccation amount varies through the size of the planetesimal itself. 
\cite{lichtenbergWaterBudgetDichotomy2019} notes that there is a strong dependence on planetesimal radius and desiccation, with larger planetesimals undergoing greater desiccation due to the significantly greater mass of \al{}.
Planetesimals were varied from \SI{1}{km} to \SI{100}{km} with 1 dex spacing. 
The core-to-planetesimal radius ratio was maintained at $\Psi = 0.5$ and \fe{} enrichment was varied between $1 \leq \Lambda\rms{60Fe} \leq 10^4$ with $1/3$ dex spacing between each simulation.
\al{} enrichment was not included, in order to focus on the influence of \fe{} on water content.

\begin{figure}
  \centering
  \includegraphics[scale=0.7]{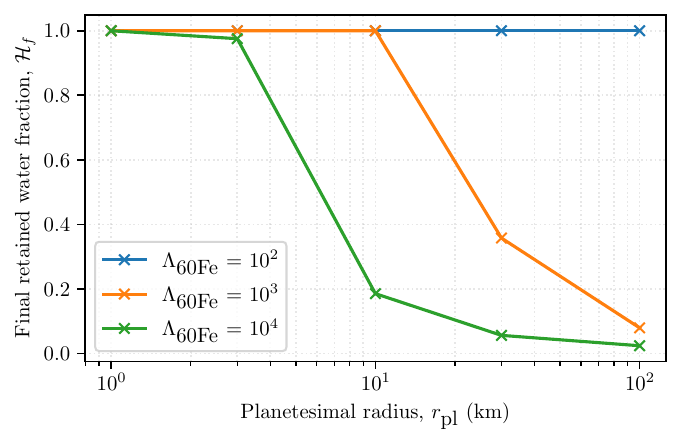}
  \caption{Final retained \water{} fraction, $\waterfracHf$ for simulations with varying planetesimal radius, $r\rms{pl}$. The \emph{core model} is utilised with a core-to-planetesimal radius ratio of $\Psi = 0.5$. There is a strong inverse correlation between $r\rms{pl}$ and $\waterfracHf$, similar to what was observed in \citet{lichtenbergWaterBudgetDichotomy2019} with water fraction, radius and \al{} enrichment, though with significantly less pronounced desiccation.}
  \label{fig:radius-comparison}
\end{figure}

Fig. \ref{fig:radius-comparison} shows that desiccation is strongly dependent on planetesimal radius.
Small planetesimals $<\SI{10}{km}$ need an extremely high degree of \fe{} enrichment in order to undergo significant desiccation, while above \SI{10}{km} there is a rapid increase in desiccation amount, which tapers out by \SI{100}{km} as water becomes increasingly rarefied.
Even in the case of large planetesimals high \fe{} enrichment is still required for desiccation; this is a recurring theme in later simulation subsets.
Lines for the $\Lambda\rms{60Fe} = 1$ and $\Lambda\rms{60Fe} = 10$ simulation subsets were not included, as no desiccation was observed in any of these simulations.

In order to constrain the parameter space of our simulations, we use a common planetesimal sizes of \SI{50}{km} and \SI{100}{km} for all subsequent simulations, which is approximately at the peak of the birth size distribution of planetesimals generated by the streaming instability mechanism \citep{2019ApJ...885...69L,2022arXiv221204509S}.

\subsubsection{Isotopic enrichment}
\label{sec:core-isotopic}

\begin{figure}
  \centering
  \includegraphics[scale=0.7]{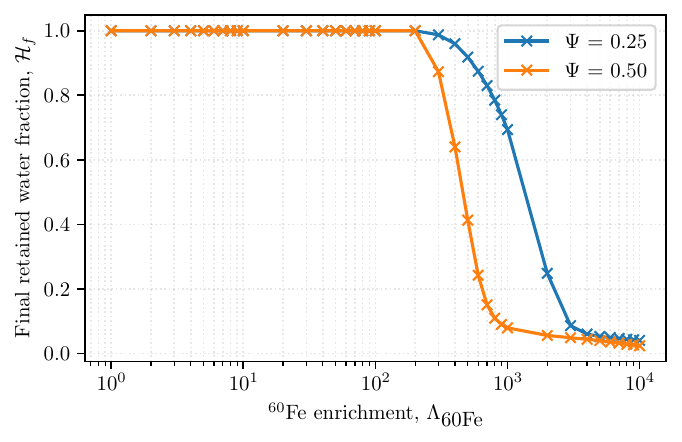}
  \caption{Final retained \water{} fraction for \emph{core model} simulations undergoing radiogenic heating solely from \fe{}. \fe{} must be enriched to greater than approximately $3 \times 10^2$ times the lower limit of the Solar System abundance for desiccation to occur.}
  \label{fig:pure-iron}
\end{figure}

Once planetesimal geometry had been constrained, we conducted an additional set of simulations in order to detail the influence of isotopic enrichment on desiccation.
Fig. \ref{fig:pure-iron} shows the results of simulations where only \fe{} enrichment occurs, $\Lambda\rms{60Fe}$ is varied from 1 to $10^4$ and $\Psi$ is varied between 0.25 and 0.5.
The results are consistent with the previous simulations, and shows that desiccation becomes significant above $200$ times the lower limit for Solar System \fe{} enrichment (recall that the lower limit is $\Lambda_{\rm 60Fe} = 1.15 \times 10^{-8}$), with a larger core resulting in more desiccation.

To compare the influence of \fe{} and \al{} on water content, a number of simulations with both SLRs at varying degrees of enrichment were conducted.
Similar to the previous simulations, two subsets were run, one where $\Psi = 0.25$ and another where $\Psi = 0.50$, \fe{} abundance was varied from $1$ to $10^3$ times Solar System enrichment in a sequence of the form $\{10^k\}$, planetesimal radius was not varied, and set as \SI{50}{km}.
Meanwhile, \al{} enrichment was restricted to 0, 1, 1.78, 5.62 and 10 times Solar System enrichment, a sequence of the form $\{10^{k/4}\}$.

\begin{figure}
  \centering
  \includegraphics[scale=0.7]{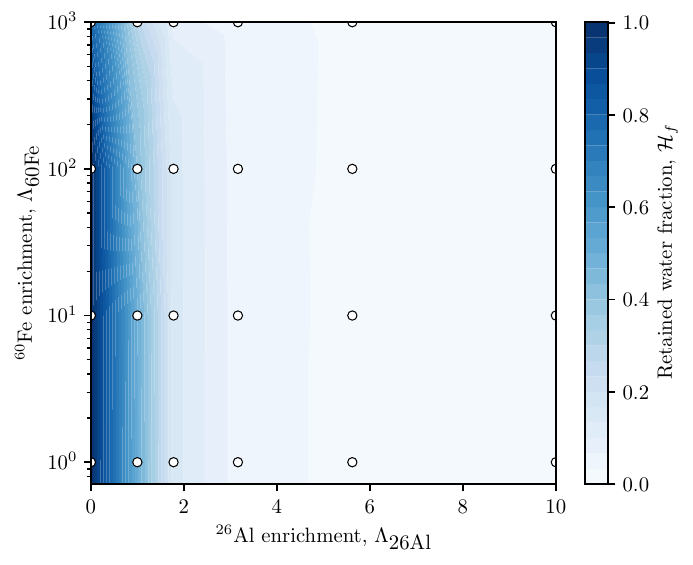}
  \caption{Comparison of final \water{} fraction, $\waterfracHf$, for simulations with a core ratio, $\Psi$, of 0.25 in the \emph{core model}. Desiccation increases dramatically as \al{} enrichment increases; however \fe{} enrichment only becomes significant with greater than approximately $3 \times 10^2 \, Z\rms{60Fe,ss,low}$ or higher.}
  \label{fig:core-surface-des}
\end{figure}

\begin{figure}
  \centering
  \includegraphics[scale=0.7]{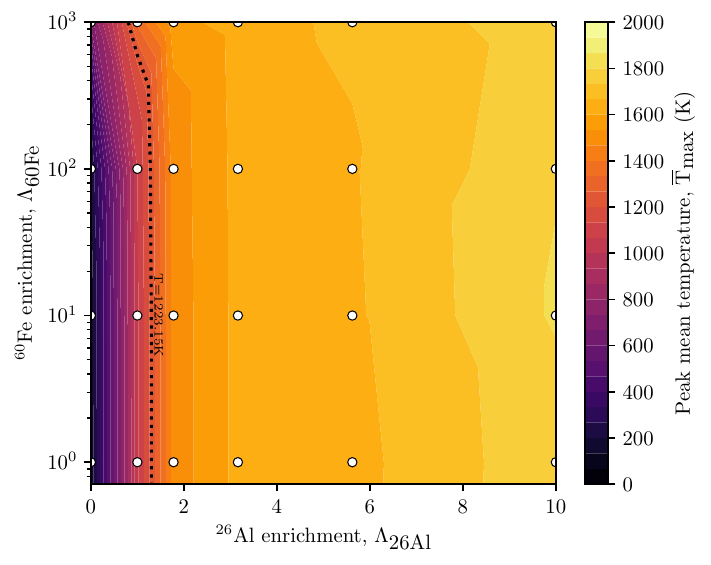}
  \caption{Comparison of peak planetesimal mean temperature, $\bar{T}\rms{peak,pl}$, for simulations where $\Psi = 0.25$ in the \emph{core model}. Similarly to Fig. \ref{fig:core-surface-des} we see that there is a strong dependence on planetesimal temperature and \al{} enrichment but an extremely weak dependence on \fe{} enrichment and planetesimal temperature. This shows that in the case of the \emph{core model} \fe{} has a weak dependence.}
  \label{fig:core-surface-temp}
\end{figure}

Fig. \ref{fig:core-surface-des} is a contour plot of the final retained water fraction, $\waterfracHf$, for the $\Psi = 0.25$ simulation set.
These plots use a continuous gradient interpolated between these simulations using Delauney triangulation, which can be used to generate contours between sparse, non-linearly spaced data.
We find that \al{} is very effective at reducing the final water content of a planetesimal, with near total desiccation happening in planetesimals with only marginally greater than Solar System \al{} abundance.
Conversely, \fe{} does have an impact, but only for simulation with little to no \al{}.
In the case of a simulation with total \al{} depletion and $\Lambda\rms{60Fe} = 1000$ a final water fraction of $\waterfracHf = 0.796$ is found, which is higher than the Solar System enrichment estimation simulation, which reports a value of $\waterfracHf = 0.509$.
Fig. \ref{fig:core-surface-temp} details the temperature change due to \al{} and \fe{} enrichment.
An identical change is observed, with \fe{} enrichment having a minimum impact on the peak mean planetesimal temperature, while \al{} enrichment rapidly pushes the planetesimal temperature through the vaporisation point.

\begin{figure}
  \centering
  \includegraphics[scale=0.70]{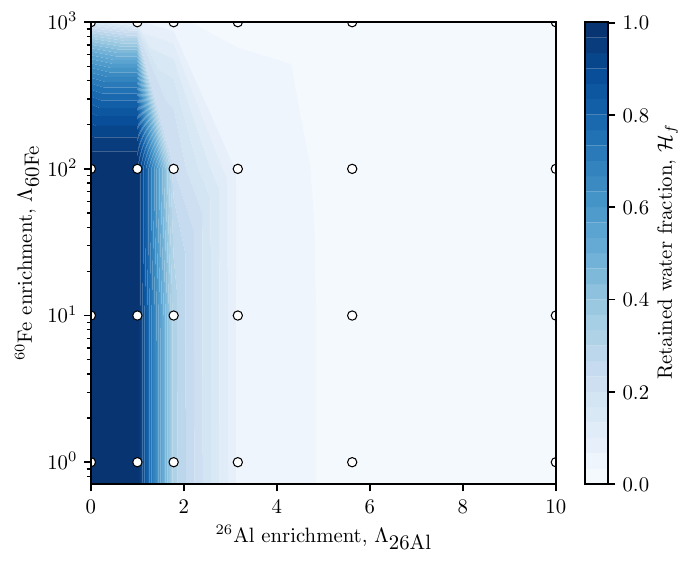}
  \caption{Comparison of $\waterfracfinal$ for \emph{core model} planetesimals where $\Psi = 0.50$, as there is a significantly lower mass of Al, there is a lower dependence on \al{} enrichment and a greater \fe{} enrichment dependence compared to Fig. \ref{fig:core-surface-des}. However, it is still not significant, as \fe{} enrichment must be 2 orders of magnitude higher than Solar System estimates to have any reasonable impact.}
  \label{fig:core-0.5-des}
\end{figure}

Fig. \ref{fig:core-0.5-des} is similar to Fig. \ref{fig:core-surface-des} for the $\Psi=0.5$ simulation set, a similar lack of influence due to \fe{} enrichment is observed.
However, low-\al{}-enrichment simulations do show markedly less desiccation, primarily due to having $\sim 30\%$ of the hydrous silicate mass of the $\Psi = 0.25$ planetesimals.
Conversely, as there is significantly more \fe{}, we observe a final water abundance fraction for the $Z_{\text{26Al}} = 0$, $\Lambda\rms{60Fe} = 1000$ simulation of $\waterfracHf = 0.153$, far below the maximum highest \fe{}-only desiccation amount of the $\Psi = 0.25$ simulations.

To summarise,\al{} is typically the primary radiogenic heating source for planetesimals with a clearly differentiated core, requiring \fe{} enrichment far greater than the lower Solar System estimates to result in significant \water{} outgassing. However, in planetary systems that are highly enriched in \fe{}, for example by supernovae enrichment, \fe{} can start to significantly contribute to planetesimal devolatilization.

\subsection{Grain model}
\label{sec:grainmodelresults}

For the second set of simulations we implemented the \emph{grain model} instead of the \emph{core model}.
Two major parameters are varied over these simulations: \fe{} enrichment $\Lambda\rms{60Fe}$ and iron volume fraction $\Phi$. \al{} enrichment $\Lambda\rms{26Al}$ is also varied, though only between 0, 1 and 10.
The parameter space of \al{} enrichment was significantly compressed compared to the \emph{core model} simulation set, in order to accommodate the expanded \fe{} enrichment space, and introduce the $\Phi$ parameter without a drastic increase in simulation count.
$\Lambda\rms{60Fe}$ was varied from 1 to $10^4$ -- with 1 dex steps -- while $\Phi$ was varied between 0.01 and 0.9.
Planetesimal radius was increased from the \emph{core model} simulation set to \SI{100}{km}.
The results of these simulations are very similar to the results of the \emph{core model}, \fe{} enrichment has to be significantly enriched compared to the Solar System in order for desiccation to occur in the case of \al{} depletion.
Meanwhile, \al{} enrichment relative to canonical Solar System levels results in a greater deal of desiccation, and enrichment beyond that results in near-total desiccation.
A divergence from the previous simulation set arises from varying $\Phi$, where high values impede desiccation from \al{}, as there is significantly less \al{} available to heat the planetesimal.
This is similar to the results observed for the \emph{core model} $\Psi = 0.5$ subset as shown in Fig. \ref{fig:core-0.5-des}.

\begin{figure*}
  \centering
  \includegraphics[scale=0.70]{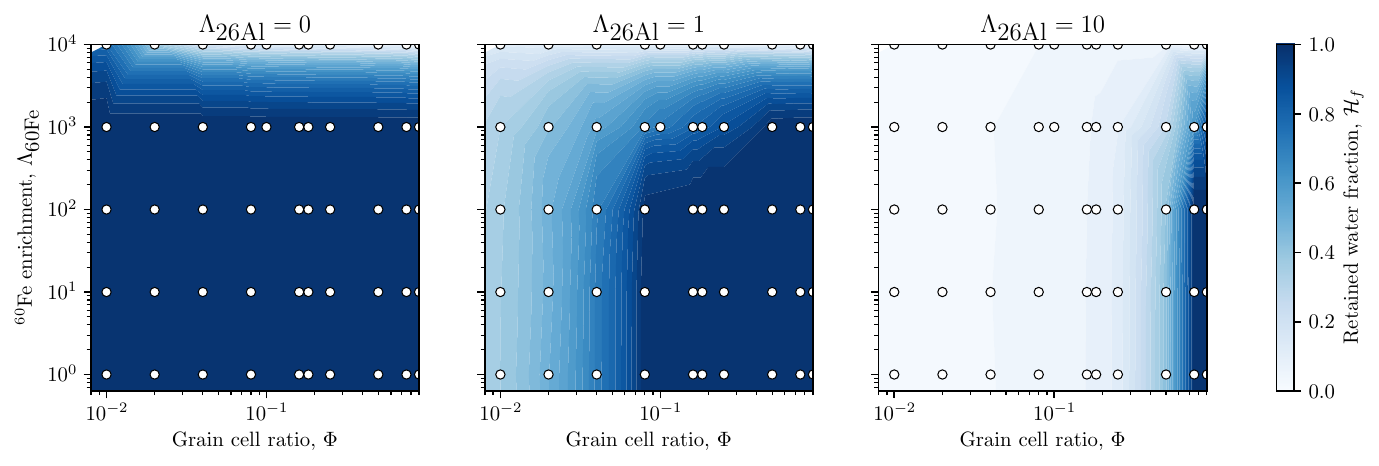}
  \caption{\emph{Grain model} comparison of $\waterfracHf$ for simulations varying \al{} enrichment, \fe{} enrichment and Fe grain volume ratio, $\Phi$. Desiccation is similarly dependent on \al{} enrichment as in the case of the \emph{core model}, while having a similar weak dependence on \fe{} enrichment. In order for desiccation to begin the planetesimal requires a large amount of iron grains by volume and an extremely large amount of enrichment ($> 10^3 \cdot Z\rms{Fe,ss}$). This suggests that even with a more physically accurate distribution of iron in a planetesimal it is still unlikely that \fe{} is significant as a radiogenic heating source in planetary formation.}
  \label{fig:grain-model-phi-des}
\end{figure*}

\begin{figure*}
  \centering
  \includegraphics[scale=0.70]{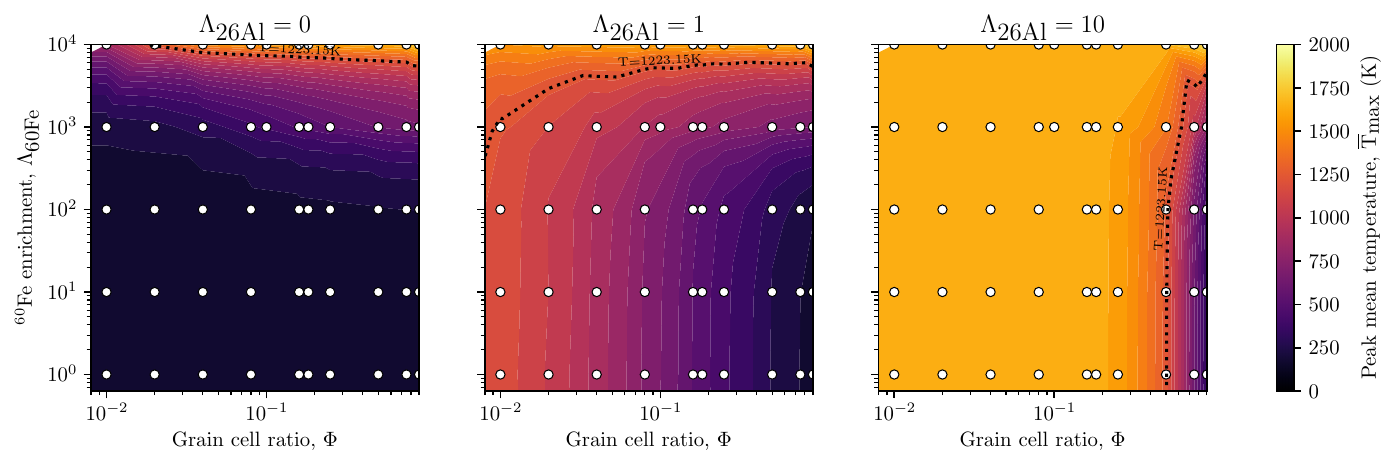}
  \caption{\emph{Grain model} comparison of peak mean planetesimal temperature for simulations varying \al{} enrichment, \fe{} enrichment and Fe grain volume ratio, $\Phi$. We note that cooling decreases in the case of \al{} enriched simulations with large quantities of low enriched Fe grains can actually lower planetesimal temperature and hinder heating and hence desiccation, as there is a lower mass of \al{} to heat the planetesimal.}
  \label{fig:grain-model-phi-temp}
\end{figure*}

Fig. \ref{fig:grain-model-phi-des} shows $\waterfracHf$ for varying values of enrichment parameters $\Lambda\rms{26Al}$, $\Lambda\rms{60Fe}$ and grain volume fraction, $\Phi$.
Data is interpolated between simulations in the same manner as Section \ref{sec:core-isotopic}.
Similarly to the \emph{core model} simulations, \fe{} enrichment has a significantly lower impact on final water content than \al{} enrichment.
In fact, the link between desiccation and \fe{} enrichment is less pronounced than with the \emph{core model}.
In the case of fully \al{}-depleted simulations, we find that no appreciable desiccation occurs until $\Lambda\rms{60Fe} > 1000$ even for simulations where $\Phi = 0.9$.
Fig. \ref{fig:grain-model-phi-temp} shows the associated temperature of these simulations, here we can see similar dependencies as with Fig. \ref{fig:grain-model-phi-des}, and that $\bar{T}\rms{peak}$ is not exceeded outside of simulations with \al{} enrichment.
In summary, the \emph{grain model} produces similar results to the \emph{core model}, as such we can infer that bulk heating is more important than heating specific areas of the planetesimal.
Furthermore, this lends further evidence that \al{} enrichment is significantly more influential with \fe{}/\al{} fractionation factors close to the Solar System, as \fe{} enrichment must be multiple orders of magnitude higher to produce any significant effect, while a similar effect occurs with a single-digit factor increase in \al{} enrichment.

\section{Discussion}
\label{sec:discuss}

\subsection{\fe{} versus \al{}}

Our simulation results suggest that radiogenic heating from \al{} is significantly more important compared to \fe{} for \fe{}/\al{} fractions close to the Solar System value.
In certain cases, with \fe{} enrichment more than $10^2$ to $10^3$ relative to the Solar System, \fe{} heating can result in significant desiccation of a planetesimal, though these values may only be realised in massive star-forming regions, or with distinct supernovae enrichment in individual systems.
Future considerations for determining which SLRs are important for the process of planetary formation would need to narrow down a typical galactic enrichment range, and whether such \fe{} enrichment levels are common on a planetary system level. Additional simulations can narrow down this parameter space further by incorporating more comprehensive models of stellar feedback on local scales \citep{2019MNRAS.485.4893N,2023MNRAS.521.4838P,2023MNRAS.525.2399P}.

\subsubsection{Other SLRs} %

Whilst other SLRs formed by stellar nucleosynthesis could be considered for simulation, such as \atom{Ca}{41} and \atom{Mn}{53} \citep{russellOriginShortLivedRadionuclides2001}, these would have a lower abundance and enrichment to \fe{}, resulting in an even lower influence on desiccation.
As such, these were not included in our simulations.

\subsection{\fe{} as a temperature sustainer?}

\begin{figure}
    \centering
    \includegraphics[width=\linewidth]{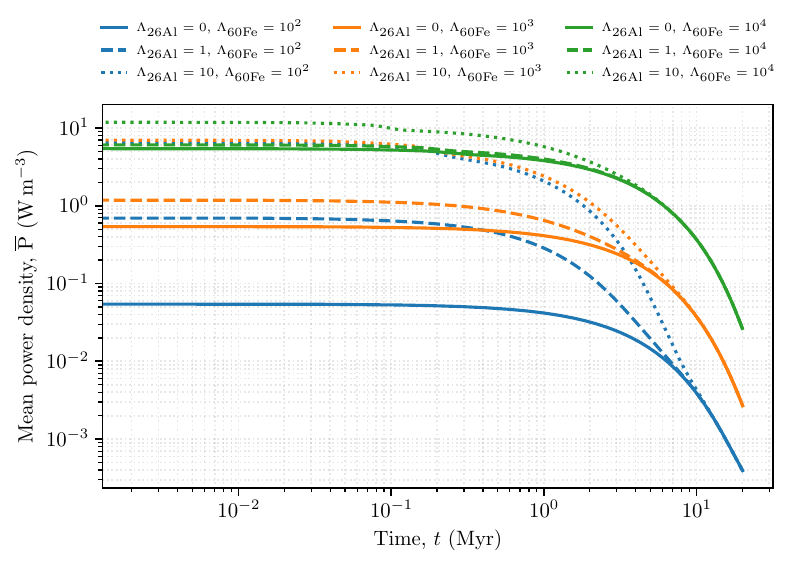}
    \caption{A comparison of mean power density, $P$, as a function of simulation time for various simulations. The \emph{grain model} is utilised where $\Phi = 0.25$, while $\Lambda\rms{60Fe}$ and $\Lambda\rms{26Al}$ are varied. In all cases specific power drops significantly between 2 and 10 Myr after the start of the simulation as \al{} decays, though  simulations with a higher content of \fe{} still have a significant power density.}
    \label{fig:comparison-time}
\end{figure}

Whilst our initial results show that there is a minimal impact of \fe{} on planetesimal heating outside highly enriched systems, the longer half-life of \fe{} could potentially sustain internal temperatures if the planetesimal was sufficiently enriched.
This can be seen in Fig. \ref{fig:heating} where the specific heating of \fe{} is greater than the specific heating rate of \al{} $\sim\SI{2}{Myr}$ from CAI formation.
In the case of Solar System abundances, however, this does not produce a significant amount of heating, and is still almost 2 orders of magnitude less than the specific heating rate of \al{} at the specific heating crossing point.
For more \fe{}-enriched planetesimals this could result in sustained heating.
A continued heating source would prevent the planetesimal from cooling and continue the vaporisation and out-gassing processes.
In order to infer the influence of \fe{} over longer time scales from our simulations, we calculated the mean averaged radiogenic heating power density:

\begin{equation}
    \overline{P} = \frac{\sum{Q\rms{T,cell}(t) / \rho\rms{cell}}}{N\rms{cells}} , 
\end{equation}

\noindent
over the planetesimal.
Fig. \ref{fig:comparison-time} shows a comparison of $\overline{P}$ over time between simulations utilising the \emph{grain model} where $\Phi = 0.25$ with varying isotopic enrichment.
After $2\,$Myr the average power density falls in the case of all simulations, becoming similar for simulations with matching \fe{} enrichment amounts.
It is clear that in all cases with significant \fe{} enrichment that \fe{} heating becomes the strongest heating mechanism near the end of the simulations, and should retain adequate heating for some time after the simulations have concluded.
Whilst this enduring heating would not change the results with our desiccation model, more complex thermochemical models with out-gassing and without a discrete evaporation temperature would lead to greater desiccation rates. 

The sustenance of high temperatures is relevant for late-formed planetesimals and cometesimals in extrasolar systems. In the Solar System, there is evidence for prolonged planetesimal formation, essentially until the very end of the disk phase \citep{2020SSRv..216...55K,2022arXiv221204509S,2023ASPC..534..907L}. Recent evidence from JWST for water-enriched inner disks in both low-mass and high-mass star-forming regions \citep{2023Natur.620..516P,2023arXiv231011074R} illustrates that desiccation of planetesimals until very late stages of the disk could impact the final composition of rocky and terrestrial exoplanets. How effective the sustenance by \fe{} is will depend on the late-stage pebble flux and dislocation of vapour from ice-rock mixtures.
Our simulations show that a very high \fe{} enrichment relative to the solar system produces temperatures amenable for devolatilization, which would be sustained until the very late stages of disk evolution.
During the late stages of the disk the varied composition of comet-like objects can impact the thermal evolution of still forming bodies within the disk (see Section \ref{sec:assumptions}) \citep{2021Icar..36314437G,2024arXiv240100231A}.
This effect is not taken into account in the simulations discussed in this paper.
As such, further investigation of the conditions where \fe{} can still maintain devolatilization in planetary systems should be performed.
Such investigations should ideally have a focus on smaller objects, high ice fractions, and varying composition (such as a varied C/O ratio) \citep{2021MNRAS.505.5654D,2021ApJ...913L..20L}.

\subsection{Possible influence on exoplanet populations}

Gaining information on the SLR distribution across planet-forming systems has been a continuous challenge, which complicates assessing their influence on exoplanet formation \citep{2018PrPNP.102....1L,parkerBirthEnvironmentPlanetary2020}. However, it has been suggested that planetary debris in polluted white dwarf systems may serve as additional constraint on the SLR distribution across planetary systems if the effects of accretion energy and SLR heating can be distinguished \citep{2013ApJ...775L..41J,2014AREPS..42...45J}. With increasing measurement precision and numerical modelling, research has shown that polluted white dwarfs indeed seem to conserve the footprint of planetesimal differentiation across a significant fraction of planetary systems \citep{2020MNRAS.492.2683B,2023NatAs...7...39B,2022MNRAS.515..395C}. 

From a star formation perspective, for a typical Initial Mass Function \citep[e.g.][]{2013MNRAS.429.1725M}, massive stars that would produce $^{60}$Fe are produced in significant numbers (i.e.\,\, $>$5) in regions that also form more than 1000 low-mass stars. Notable examples of such regions in our Galaxy are Cyg OB2 \citep[e.g.][]{2015MNRAS.449..741W}, Westerlund 1 \citep[e.g.][]{2005A&A...434..949C} and the Arches cluster \citep{2002ApJ...581..258F}. However, stars likely form stochastically, and in this scenario the only minimum limit on the most massive stars that can form in a region is the total mass of the region itself \citep{2006ApJ...648..572E}, though see \citet{2006MNRAS.365.1333W}. However, in any region that produces massive stars, significant photoevaporation of the gaseous component of protoplanetary discs occurs \citep[e.g.][]{2001MNRAS.325..449S,2004ApJ...611..360A,2019MNRAS.485.4893N,2019MNRAS.490.5678C,2022EPJP..137.1132W}, meaning that the planetary systems that could be heavily enriched in $^{60}$Fe would likely be devoid of gas giants (Patel et al, submitted). 

An alternative enrichment scenario, which avoids the destructive photoevaporation from massive stars, is from Asymptotic Giant Branch stars \citep[e.g.][]{2016ApJ...825...26K,2018PrPNP.102....1L,parkerIsotopicEnrichmentPlanetary2023}. AGB stars are a stellar evolutionary phase that all 1 -- 8\,M$_\odot$ stars undergo \citep{2005ARA&A..43..435H}, and these stars produce significant yields of \fe{} and \al{} via their winds \citep{2018MNRAS.475.2282V}. AGB stars were previously discounted as a viable source of SLR enrichment due to the supposed low probability of an evolved star encountering a young star as it forms its planetary system \citep{1994ApJ...421..605K}. However, \cite{parkerIsotopicEnrichmentPlanetary2023} report the serendipitous Gaia DR3 discovery of an AGB star interloping through the young star-forming region NGC\,2264, and furthermore, show that under reasonable assumptions for the initial conditions of the star-forming region, an AGB star could enrich the Solar System in both \fe{} and \al{}, with some stars attaining higher \fe{} ratios than in the Solar system (e.g. $\Lambda_{\text{60Fe}} \sim 10^{-4}$).

In addition, studies of planet-forming disks indicate that planetesimal formation is rapid, typically operating on a timescale of $10^5$ yr \citep{Drazkowska2022,2022arXiv220309930M}. Combined, this suggests that a significant fraction of exoplanetary systems are expected to undergo significant SLR-driven heating during the formation of their planets, with substantial fractionation possible between \fe{} and \al{}.

Exoplanet surveys have just started to reach into the low-mass regime, where it is possible to distinguish bulk-level under- and over-densities in the size regime of super-Earths and below \citep{Wordsworth2022,2023arXiv230610100P}. Further investigation of these population trends will constrain the effects of accretion environment and geophysical evolution on bulk composition. While typically astronomical studies tend to focus on atmospheric escape and disk migration as the main effects on volatile bulk composition \citep{2020SSRv..216...86V,2021JGRE..12606639B}, we have shown here that variations in the main SLRs can drive substantial devolatilization of the planetesimal building blocks. From a Solar System perspective, meteoritic evidence of initial incorporation of water, and other volatiles and rapid devolatilization of the building blocks of Earth and its planetesimal precursors is accumulating \citep{2017RSPTA.37560209S,2019E&PSL.52615771M,2021Sci...371..365L,2021PNAS..11826779H,2021NatAs...5..356G,2023GeCoA.340..141P,2022GeCoA.316..201L,2023Natur.615..854N,2023E&PSL.61518202S,grant2023bulk}.
This presents a challenge for a wide spread of potentially Earth-like rocky planets with a similar geodynamic an climatic regime, as the mass fraction of surface water to allow for the coexistence of oceans and exposed land with a geodynamic regime as the Earth is limited to within a relatively narrow range \citep{2014ApJ...781...27C,2015ApJ...801...40S,2017SSRv..212..877N}. Observational insights on the water content and devolatilization trends of planetary debris around polluted white dwarfs \citep{2013Sci...342..218F,2022MNRAS.515..395C} and in debris disks \citep{2021ApJ...913L..20L,2022arXiv220203053M,2023arXiv230701574B} will be crucial to narrow down the potential spread in these effects. 

Ultimately, increased understanding of the influence of planetesimal degassing is required to understand the information received from exoplanet population analyses, planetary debris, and detailed characterization of individual low-mass exoplanets.

\section{Conclusions}
\label{sec:conclusion}

Overall we conclude that the influence of radiogenic heating on water content in protoplanetary systems due to \fe{} is weaker compared to heating from the typically more abundant and energetic \al{}.
However, the contribution of \fe{} to the devolatilization of early-formed planetesimals becomes substantial for enrichment levels of $^{60}\mathrm{Fe}/^{56}\mathrm{Fe} \gtrsim 10^{-6}$, which is the upper end range of enrichment values that has previously been inferred for the Solar System.
In exoplanet systems that formed in high-mass star-forming regions with substantial supernovae feedback, \fe{} may thus be an important contributor to SLR-driven heating.
Our parameter space explored \al{} enrichment over an order of magnitude, while \fe{} was explored over four orders of magnitude, we found that \fe{} did not become dominant unless the system was \al{} depleted, \fe{} enriched, or possessed a high iron content. Each of these characteristics may be fullfilled in exoplanetary systems forming in diverse star-forming regions, however, we anticipate \al{} to be the main driving factor of planetesimal differentiation and devolatilization. 

Fractionation in \fe{} and \al{} as a function of the birth star-forming region is expected due to intrinsic stochasticity of star formation \citep[where variations in the IMF lead to significantly different numbers of massive stars, and therefore enrichment, e.g.][]{2017MNRAS.464.4318N}, the stochasticity of dynamical evolution of young star-forming regions, which affects the amount of SLR ejecta that may be captured by an individual planetary system (Patel et al., submitted) and the short half-lives of the main SLRs. Our results thus suggest that SLR-driven internal desiccation of planetesimals may contribute to compositional scatter across exoplanetary systems, following the initial SLR enrichment of exoplanetary systems at their birth. Future work shall explore the geochemical effect of water flow and iron oxidation inside volatile-rich planetesimals, and the compositional effects of ongoing accretion on the final composition and atmospheric diversity of low-mass exoplanets.

\section*{Acknowledgements}

JWE and RJP acknowledge support from the Royal Society from a Dorothy Hodgkin Fellowship and an Enhancement Award. TL was suppported by a grant from the Branco Weiss Foundation.

\section*{Data availability}

The data underlying this article is stored at The University of Sheffield ORDA repository.
The data and scripts to reproduce the figures included in this paper are stored as well as the initial conditions for each simulation, such that the simulations can be performed with a suitable workstation.
The repository is located at
\weblink{https://doi.org/10.15131/shef.data.23815599.v1}.

\bibliographystyle{mnras}
\bibliography{references,tim-merged-refs}

\bsp	%
\label{lastpage}
\end{document}